\documentclass{webofc}

\usepackage[varg]{txfonts}   
\usepackage{hyperref}
\usepackage{url}
\hypersetup{colorlinks=true,citecolor=blue,urlcolor=blue,linkcolor=blue}
\begin{document}
\title{Fast timing silicon R$\&$D for the future Electron-Ion Collider}
%
%

\author{\firstname{Xuan} \lastname{Li}\inst{1}\fnsep\thanks{\email{xuanli@lanl.gov}} \and
     \firstname{Eric} \lastname{Renner}\inst{1} \and  
     \firstname{Ming} \lastname{Liu}\inst{1} \and
     \firstname{Walter} \lastname{Sondheim}\inst{1} \and
        \firstname{Carlos} \lastname{Solans Sanchez}\inst{2} \and
         \firstname{Marcos} \lastname{Vazquez Nuñez}\inst{3} \and
    \firstname{Vicente} \lastname{Gonzalez}\inst{3} \and        
        \firstname{Yasser} \lastname{Corrales Morales}\inst{4}
}

\institute{Los Alamos National Laboratory, USA 
\and
           CERN, Switzerland
\and
           University of Valencia, Spain
\and
          Massachusetts Institute of Technology, USA           
          }

\abstract{The proposed Electron-Ion Collider (EIC) will utilize high-luminosity high-energy electron+proton ($e+p$) and electron+nucleus ($e+A$) collisions to solve several fundamental questions including searching for gluon saturation and studying the proton/nuclear structure. Complementary to the ongoing EIC project detector technical prototype carried out by the ePIC collaboration, a Depleted Monolithic Active Pixel Sensor (i.e., MALTA2) based fast timing silicon tracking detector (FMT) has been proposed to provide additional hits for track reconstruction in the forward region at the EIC to improve the overall track reconstruction quality. The fast timing resolution of the MALTA2 technology will help reject background events at the EIC as well. Progress of latest MALTA2 R$\&$D, the development of a new MALTA2 quad-sensor prototype module and impacts of the proposed FMT in EIC physics studies will be discussed.
}
\maketitle
\section{Introduction}
\label{sec-intro}
The future Electron-Ion Collider (EIC) is expected to start construction at Brookhaven National Laboratory in 2025. It will support two interaction points (IP6 and IP8), where the detector located at IP6 is referred to as the EIC project detector and the one at IP8 is named as detector II. A series of electron+proton ($e+p$) and electron+nucleus ($e+A$) collisions (A=2-238) at the center of mass energies of 20-141 GeV will be operated at the EIC \cite{eic_YR}. The bunch crossing period at the EIC is around 10.2~ns with a 25~mrad beam crossing angle is introduced to IP6 to reduce impacts from beam backgrounds. The current technical design of the EIC project detector at IP6 led by the ePIC collaboration utilizes the 65~nm Monolithic Active Pixel Sensor (MAPS) \cite{its3}, the Micro Pattern Gaseous Detector (MPGD) \cite{mpgd} and the AC-coupled Low Gain Avalanche Diode (AC-LGAD) \cite{ac-lgad} to form its vertex and tracking subsystem. In addition to ongoing EIC detector development, a Depleted MAPS (i.e., MALTA2 \cite{malta2}) based Fast MAPS Tracker (FMT) had been proposed as a new subsystem for either the ePIC detector upgrade or the EIC detector II. The design and R$\&$D for the proposed FMT is underway supported by the EIC generic R$\&$D program.
\section{MALTA2 R$\&$D progress}
\label{sec-tech}
\begin{figure}[h]
\centering
\includegraphics[width=0.9\textwidth,clip]{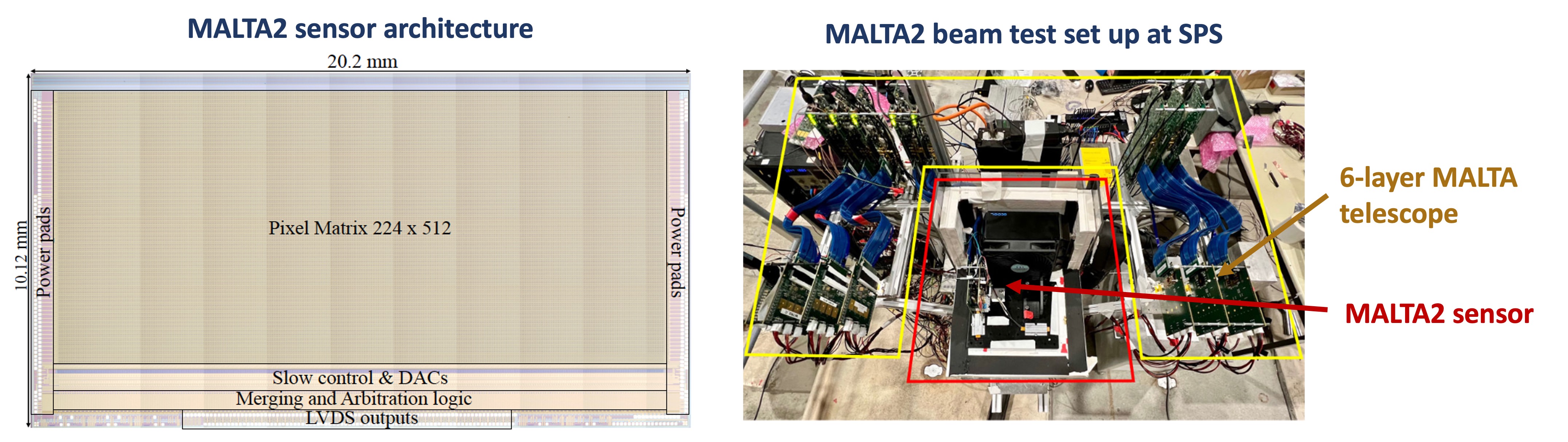}
\caption{The architecture of the MALTA2 sensor (left) and the CERN SPS beam test setup to characterize the MALTA2 sensor (right).}
\label{fig-1}       
\end{figure}

\begin{figure}[h]
\centering
\includegraphics[width=0.99\textwidth,clip]{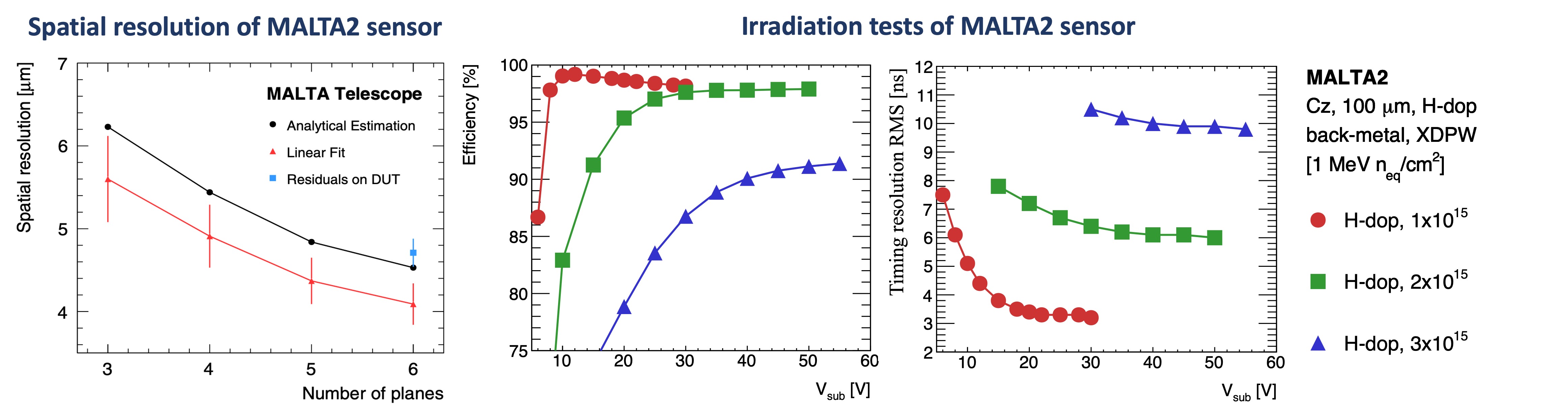}
\caption{Spatial resolution of the MALTA2 sensor from the CERN SPS beam tests \cite{malta2_perform} (left). The bias voltage dependent efficiency (middle) and timing RMS (right) of the MALTA2 sensor with the irradiation dose at 1-3 $\times$ 10$^{15}$~1MeV $n_{eq}$/cm$^{2}$ and 100 $\mu$m wafer thickness \cite{malta2_rad}.}
\label{fig-2}       
\end{figure}

The MALTA2 sensor fabricated with the Tower 180~nm CMOS technology consists of a 224 by 512 pixel array on a 1~cm by 2~cm active area (see the left panel of figure~\ref{fig-1}) \cite{malta2}. The average digital power consumption of the MALTA2 sensor is around 10~mW/cm$^{2}$ and its analog power consumption can reach around 70~mW/cm$^{2}$. Beam tests have been performed at different facilities to characterize the MALTA2 sensor and the right panel of Figure~\ref{fig-1} shows the test setup for a MALTA2 sensor placed in the middle of a 6-layer MALTA telescope at CERN SPS. Test results of MALTA2 sensor with 100~$\mu$m thickness \cite{malta2_perform} are shown in Figure~\ref{fig-2}. The average hit spatial resolution is around 4.1~$\mu$m and the timing resolution is around 2.1~ns for individual MALTA2 sensors \cite{malta2_perform}. Minimal performance loss has been found on MALTA2 sensors at the irradiation dose around 10$^{15}$~1MeV $n_{eq}$/cm$^{2}$ \cite{malta2_rad}, which is significantly higher than the expected irradiation level at the EIC.

\section{The proposed Fast MAPS Tracker (FMT) design}
\label{sec-design}
\begin{figure}[h]
\centering
\includegraphics[width=0.99\textwidth,clip]{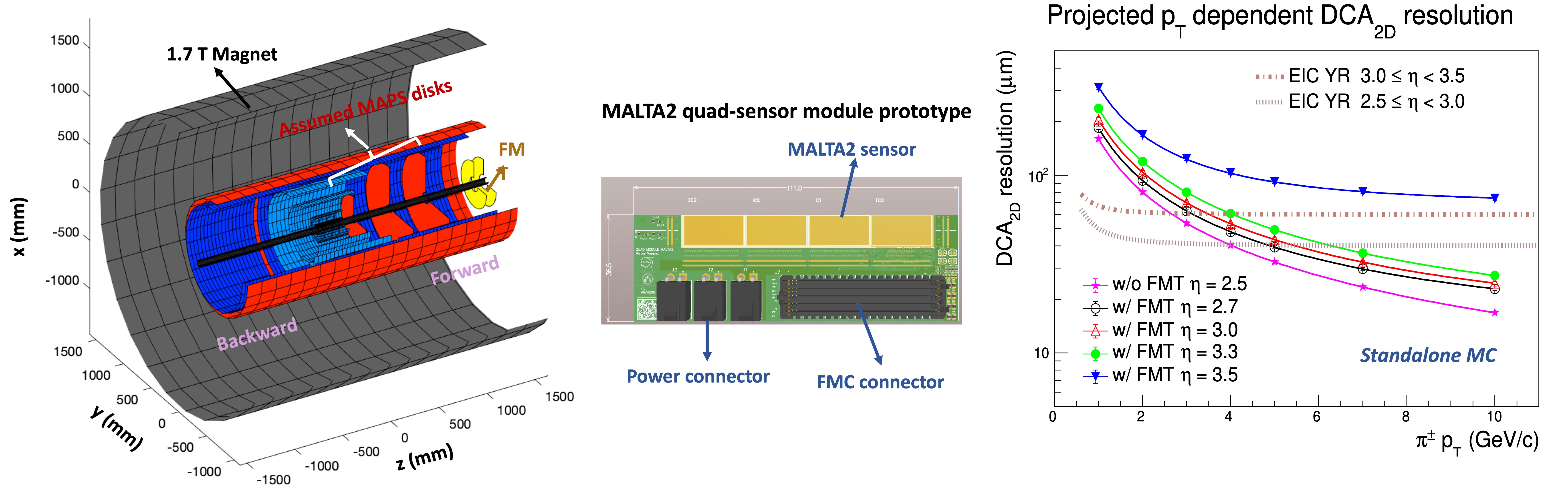}
\caption{The proposed FMT geometry implemented in fast simulation (left). The first design of the MALTA2 quad-sensor prototype module (middle). The transverse momentum dependent track $DCA_{2d}$ resolution in different pseudorapidity regions with FMT evaluated in standalone simulation (right).}
\label{fig-3}       
\end{figure}

The MALTA2 based Fast MAPS Tracker (FMT) is proposed to consist of 2 disks in the forward region ($\eta > 2.5$) near the edge of the EIC central magnet. Another candidate location is to place the FMT disks in the backward region ($\eta < -2.5$). As a starting point, the EIC IP6 beam pipe and three 65~nm MAPS (the ePIC default technology) disks with the pseudorapidity coverage up to 2.5 have been used to develop the FMT design. Several optimization iterations have been applied for the FMT geometry and the left panel of Figure~\ref{fig-3} presents the optimized FMT geometry in the forward region and the corresponding geometry parameters are summarized in table~\ref{tab-1}. The backward geometry optimization will be performed later on.

\begin{table}
\centering
\caption{The geometry of the proposed FMT in the forward pseudorapidity region}
\label{tab-1}       
\begin{tabular}{llllll}
\hline
Disk index & Inner radius & Outer radius & z location & Material budget & Hit efficiency \\\hline
1 & 7.014~cm & 23.095~cm & 145~cm & 0.74$\%$X/X$_{0}$ & 98$\%$  \\
2 & 7.014~cm & 23.095~cm & 165~cm & 0.74$\%$X/X$_{0}$ & 98$\%$ \\\hline
\end{tabular}
\end{table}

The proposed FMT will utilize either groups of MALTA2 quad-sensor staves/modules or MALTA2 octuple-sensor staves/modules to build the disk. The middle panel of Figure~\ref{fig-3} shows the first prototype design of the MALTA2 quad-sensor module carrier board. Production of the MALTA2 quad-sensor is underway and performance characterization has been planned using lab bench and beam tests at CERN. Further development of the MALTA2 octuple-sensor module is underway. Tracking performance such as spatial and momentum resolutions with the proposed FMT geometry (see Table~\ref{tab-1}) has been evaluated in standalone single particle simulation. The right panel of Figure~\ref{fig-3} shows the track $p_{T}$ dependent transverse Distance of Closest Approach (DCA$_{\rm{2D}}$) resolution with and without the FMT in comparison with the EIC yellow report detector requirement \cite{eic_YR}. The proposed FMT can extend the track reconstruction of the 3 MAPS disks to the forward pseudorapidity region of $2.5<\eta<3.3$, and meet the EIC detector requirements for tracks with $p_{T}> 2$~GeV/c.

\section{Physics impacts of the proposed FMT}
\label{sec-phys}
\begin{figure}[h]
\centering
\includegraphics[width=0.44\textwidth,clip]{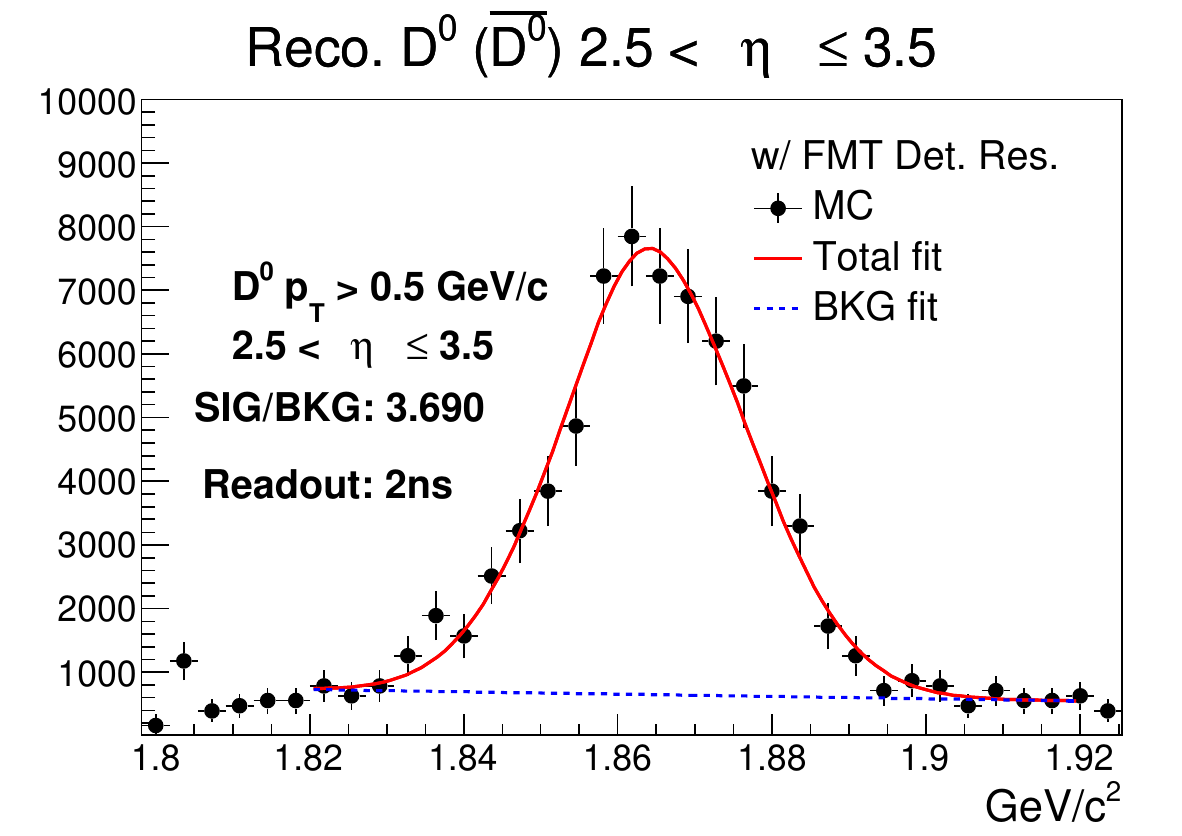}
\includegraphics[width=0.44\textwidth,clip]{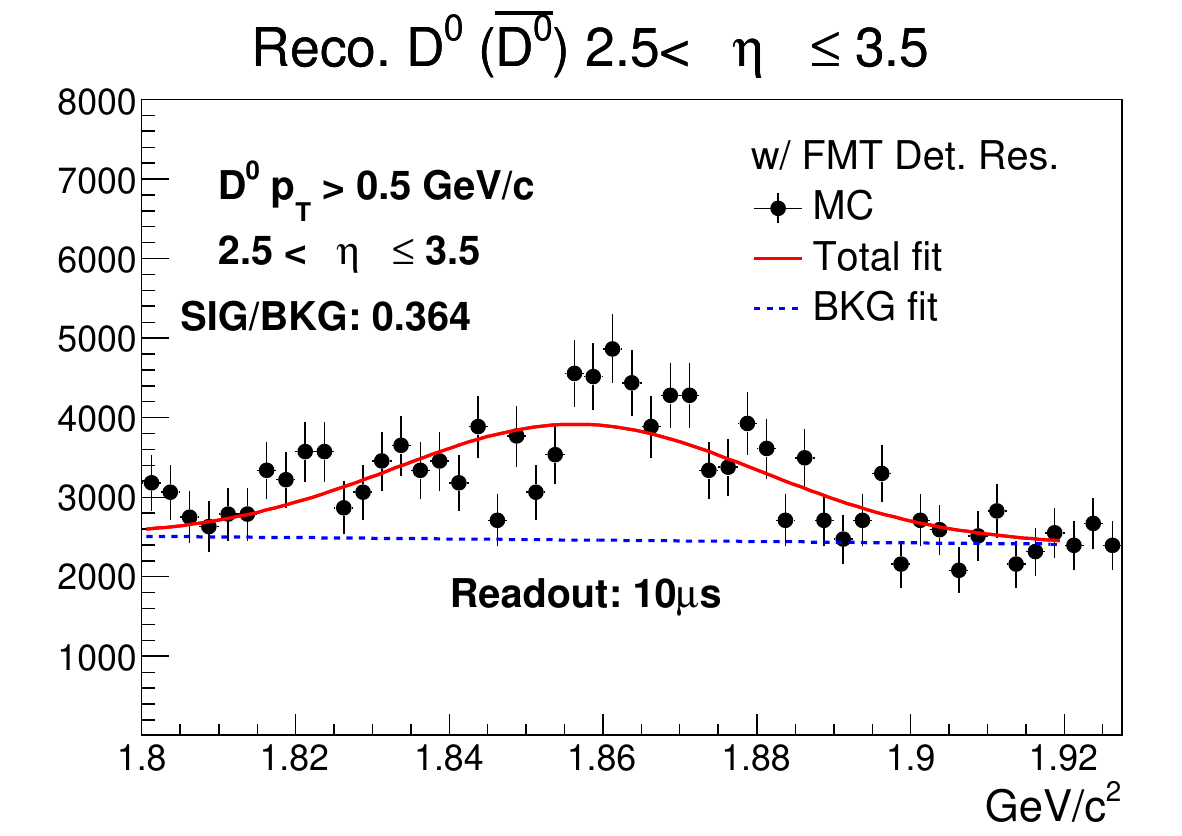}
\caption{Invariant mass spectrum of reconstructed $D^{0}$ ($\overline{D^{0}}$) in the pseudorapidity regions of $2.5~<~\eta~\le~3.5$ with 2~ns readout window (left) and with 10~$\mu$s readout window (right).}
\label{fig-4}       
\end{figure}

Fast timing is another advanced feature of the proposed FMT. Recent beam tests have demonstrated that the MALTA2 sensor can accomplish data acquisition within around 2~ns timing window \cite{malta2_perform}. The estimated collision rate at the EIC is around 500~kHz, which is equivalent to a 2~$\mu$s data collection window. The FMT will be able to carry out physics measurements for individual collisions at the EIC, which can significantly reject pileup events. A PYTHIA8 \cite{py8} based simulation framework has been setup to evaluate the impact on forward heavy flavor hadron reconstruction using different readout time windows. Figure~\ref{fig-4} compares the invariant mass of reconstructed $D^{0}$ ($\overline{D^{0}}$) ($D^{0}/\overline{D^{0}} \rightarrow K^{\mp}\pi^{\pm}$) with the projected FMT spatial resolutions in the $2.5<\eta\le3.5$ region using two different readout timing windows: 2~ns (left) and 10~$\mu$s (right). The proposed FMT with around 2~ns readout can significantly reduce the combinatorial background contributions to the forward heavy flavor reconstruction at the EIC, which has a great sensitivity in exploring the heavy quark hadronization process under different medium conditions \cite{eic-hf}.

\section{Summary and Outlook}
\label{sec-sum}
Latest bench and beam tests have demonstrated good spatial and timing resolutions for MALTA2 sensors. As evaluated in recent simulation studies, the proposed MALTA2 based FMT will help reduce pileup and combinatorial background contributions for forward heavy flavor reconstruction at the EIC. Further ongoing R$\&$D for the FMT, which includes design and production of a MALTA2 quad-sensor prototype module, will provide more technical details to validate the overall detector performance.

\end{document}